\documentclass[aps,prb,showpacs,superscriptaddress,10pt,twocolumn]{revtex4}

\usepackage{amsmath} 
\usepackage{graphicx}

\newcommand{\cI}{\mathcal{I}}
\newcommand{\cK}{\mathcal{K}}
\newcommand{\cM}{\mathcal{M}}
\newcommand{\cT}{\mathcal{T}}
\newcommand{\upd}{\mathrm{d}}
\newcommand{\sech}{\textrm{sech}}

\renewcommand{\surname}[1]{\textsc{#1}}
\makeatletter\renewcommand{\Dated@name}{Posted on the arXiv on }\makeatother

\begin{document}

\title{One-dimensional transport revisited: A simple and exact solution for phase disorder}

\date{15 April 2013}  

\author{Hui Khoon \surname{Ng}}
\email{cqtnhk@nus.edu.sg}
\affiliation{Centre for Quantum Technologies, National University of Singapore, Singapore 117543}
\affiliation{DSO National Laboratories, Singapore 118230}
\author{Berthold-Georg \surname{Englert}}
\email{cqtebg@nus.edu.sg}
\affiliation{Centre for Quantum Technologies, National University of Singapore, Singapore 117543}
\affiliation{Department of Physics, National University of Singapore, Singapore 117542}

\begin{abstract}
Disordered systems have grown in importance in the past decades, with similar phenomena \mbox{manifesting} themselves in many different physical systems. Because of the difficulty of the topic, theoretical progress has mostly emerged from numerical studies or analytical approximations. Here, we provide an exact, analytical solution to the problem of uniform phase disorder in a system of identical scatterers arranged with varying separations along a line. Relying on a relationship with Legendre functions, we demonstrate a simple approach to computing statistics of the transmission probability (or the conductance, in the language of electronic transport), and its reciprocal (or the resistance). Our formalism also gives the probability distribution of the conductance, which reveals features missing from previous approaches to the problem.
\end{abstract}


\maketitle

\section{\label{sec:Intro}Introduction}

Disorder is ubiquitous in nature. 
The issue of universal properties of disordered systems has sparked much interest, both experimental and theoretical, beginning with Anderson localization in electronic systems, and by now including phenomena observable in a wide range of physical systems.

Transport in one-dimensional (1D) disordered systems, being the simplest problem, is naturally the first to be understood. 
Studies on electronic conductivity by modeling a 1D wire as an array of potentials with random shapes and/or random spacings abound. For an excellent review of the state of affairs for 1D systems in 1982, we refer the reader to the article by Erd{\"o}s and Herndon.\cite{Erdos82} Overviews of subsequent developments can be found in Refs.~\onlinecite{Kramer93}, \onlinecite{Pendry94}, and \onlinecite{Beenakker}.

Despite the comprehensive understanding of the 1D problem, owing to the difficulty of the subject, exact analytical results are few and far between.
Perturbative treatments, typically in the limit of weak scattering, are the main tools for analytical studies: the scaling theory,\cite{Anderson80} the Born approximation for scattering,\cite{Abrikosov81} the DMPK equation,\cite{DMPK} and many more.
Exact results emerge mostly from numerical studies, or from the group-theoretic approaches of Refs.~\onlinecite{Erdos82}, \onlinecite{Kirkman84}, and \onlinecite{Pendry94}, which apply in general situations, but are nevertheless somewhat complicated because of the use of direct products of transfer matrices.

In this article, we revisit the simplest problem of a 1D single-scattering-channel system where the disorder in question is one of uniformly distributed phase disorder.
A physical realization of a system manifesting such a disorder comprises a stack of identical semi-transparent glass plates, with random spacings between adjacent plates. The task is to explore the transmission of a laser beam through the random stack. 
Because the typical separation between plates is much larger than the wavelength of the light, a uniform distribution of separation will describe the phase disorder well.
Such a system was already considered by Stokes \cite{Stokes} in 1862, who gave a ray-optical treatment.

Wave-optical investigations came much later; 
see Ref.~\onlinecite{Buchwald89} for the history of the subject.
In fact, the 1984 work by Perel' and Polyakov \cite{PP84} deals with a more
general 1D transport problem that is treated with pertinent approximations;
the situation considered here is contained as a special case, for which these
approximations are not needed.
This particular case was also investigated by Berry and Klein \cite{Berry97}
in 1997 (which work inspired our current efforts).
We employ a different strategy that exploits fully the recurrence relation
established in Ref.~\onlinecite{Lu10}, which facilitates an exact yet simple
analytical treatment. 
 
Disorder average of all moments of the conductance and resistance (that is, the transmission probability and its reciprocal) can be derived easily within a single framework.
The same framework further permits direct derivation of the exact probability distributions for the conductance and resistance, which allows for computation of all statistics of the disordered system. 
This goes beyond past work that reconstructed the distribution for the conductance from its moments,\cite{Pendry94} or derivations that relied on perturbative approaches.\cite{Gert59,Abrikosov81}
The simplicity in our solution lies in a relation with Legendre functions, whose properties are well studied and are easily amenable to analytical manipulations.
Our exact solution for this simple model can conceivably be used as the starting point for perturbation towards other more realistic systems  (see, for instance, the recent experimental proposal of Ref.~\onlinecite{Gavish05}).

The article will proceed as follows:
After setting up the problem in Section \ref{sec:Setup}, we describe (Section \ref{sec:Recur}) how to average over disorder using a recurrence relation, previously derived in Ref.~\onlinecite{Lu10}. 
This recurrence relation is applicable even for general disorder. 
Specializing to uniform phase disorder, in Sections \ref{sec:Resistance} and \ref{sec:Conductance}, we make use of a close link to Legendre functions to write down closed-form expressions for the expected values of moments of conductance and resistance. 
Comparisons with existing results in the weak-scattering, long-chain limit are presented in Section \ref{sec:WeakScat}.
The recurrence relation also gives the exact probability distributions for the conductance and the resistance, and these probability distributions are studied in detail in Section \ref{sec:Prob}. 
We close with a summary in Section \ref{sec:Conc}.

\section{\label{sec:Setup}Problem setup: Transfer-matrix description of the scattering process}
Consider a 1D chain of scatterers, such as a stack of semi-transparent glass plates, or a string of impurities. 
Both the scatterer and the scattered particle are assumed to be scalar particles, or at least only scalar degrees of freedom participate in the scattering process.
The scatterers are identical, but sit at locations with varying distances between adjacent scatterers.
The distances between adjacent scatterers are the random variables in our system---the disorder.

\begin{figure}
\begin{center}
\includegraphics[width=0.95\columnwidth]{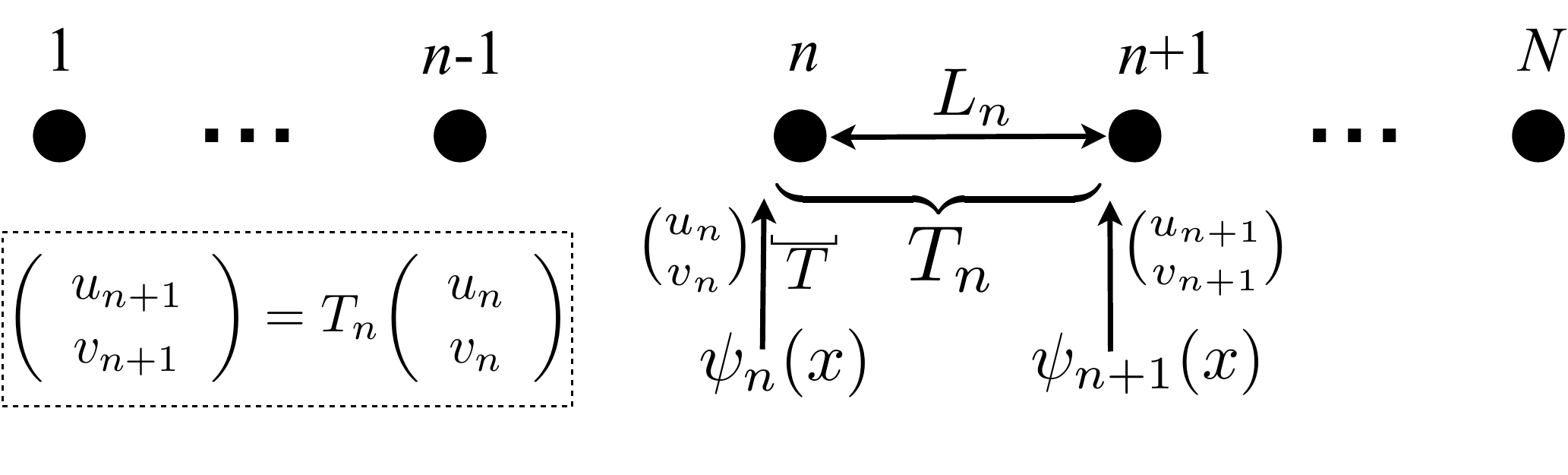}
\caption{\label{fig:FigTransfer}The transfer matrix description of the scattering process. $T$ is the transfer matrix for scattering due to a scatterer by itself (for example, a glass plate); $T_n$ includes the effects of the spacing $L_n$ between the $n$th and ($n$+1)th scatterers. $\psi_n(x)$ is the wave function of the scattered particle just before it hits the $n$th scatterer, and can be written in terms of left- and right-moving waves as $\psi_n(x)=u_ne^{\mathrm{i}kx}+v_ne^{-\mathrm{i}kx}$, where $k$ is the momentum of the scattered particle.}
\end{center}
\end{figure}

The scattering of the particle off the $n$th scatterer can be summarized by a transfer matrix $T_n$ which propagates the wave function of the scattered particle past the $n$th scatterer (see Figure \ref{fig:FigTransfer}).
$T_n$ is obtained by solving the scattering problem for the $n$th scatterer, and accounts for multiple scattering off the same scatterer due to reflected waves from adjacent scatterers. 
For our problem, $T_n$ can be decomposed into two parts,
\begin{equation}
T_n=D(kL_n)\,T\quad\textrm{with }D(\varphi)= {\left(\begin{array}{cc}\mathrm{e}^{\mathrm{i}\varphi}&0\\0&\mathrm{e}^{-\mathrm{i}\varphi}\end{array}\right)},
\end{equation}
where $T$ describes the scattering due to the $n$th scatterer itself, and $D(kL_n)$ accounts for the phase acquired by the scattered particle in traveling distance $L_n$ between the $n$th and ($n$+1)th scatterers. 
$T$ takes the general form
\renewcommand{\arraystretch}{1.15}
\begin{align}
T&=D(\beta)\,\cT(\vartheta)\,D(\alpha)\nonumber\\
\textrm{with}\quad\cT(\vartheta)&={\left(\begin{array}{cc}\cosh\bigl(\frac{1}{2}\vartheta\bigr)&\sinh\bigl(\frac{1}{2}\vartheta\bigr)\\\sinh\bigl(\frac{1}{2}\vartheta\bigr)&\cosh\bigl(\frac{1}{2}\vartheta\bigr)\end{array}\right)}.\label{eq:T1}
\end{align}
Here, $\alpha$ and $\beta$ denote overall phases, and $\cT(\vartheta)$ contains the transmission amplitude $t= \bigl[\cosh(\frac{1}{2}\vartheta)\bigr]^{-1}$ and the reflection amplitude $r=\sqrt{1-t^2}=\tanh\bigl(\frac{1}{2}\vartheta\bigr)$. 
Since the scatterers are identical, the same $T$ describes every scatterer. 
The disorder that stems from the variable separation between adjacent scatterers is characterized by $L_n$, for $n=1,\ldots, N$ (with $L_N= 0$).
\renewcommand{\arraystretch}{1.0}

The total transfer matrix, for a fixed configuration of $N$ scatterers, is
\begin{eqnarray}
&& T^{(N)}= T_NT_{N-1}\ldots T_1\\
&=&D(\beta)\,\cT(\vartheta)\,D{\Bigl(\frac{\varphi_{N-1}}{2}\Bigr)}\,\cT(\vartheta)\ldots D{\Bigl(\frac{\varphi_1}{2}\Bigr)}\,\cT(\vartheta)\,D(\alpha),\nonumber
\end{eqnarray}
where $\varphi_n= 2(\alpha+\beta+ k L_n)$.
$T^{(N)}$ can also be written in the form of Eq.~\eqref{eq:T1}, with an effective $N$-scatterer transmission amplitude $t^{(N)}= \bigl[\cosh\bigl(\frac{1}{2}\vartheta^{(N)}\bigr)\bigr]^{-1}$, so that
\begin{equation}
T^{(N)}=D\bigl(\beta^{(N)}\bigr)\,\cT\bigl(\vartheta^{(N)}\bigr)\,D\bigl(\alpha^{(N)}\bigr)
\end{equation}
with overall phases $\alpha^{(N)}$ and $\beta^{(N)}$. 
These phases and $\vartheta^{(N)}$ can be recursively expressed in terms of $\vartheta$ and the individual spacings $L_n$. 
This can be understood by adding one more scatterer to the end of the chain of $N$ scatterers. 
The total transfer matrix is now $T^{(N+1)}=T_{N+1}T^{(N)} = D\bigl(\beta^{(N+1)}\bigr)\,\cT\bigl(\vartheta^{(N+1)}\bigr)\,D\bigl(\alpha^{(N+1)}\bigr)$, with $\vartheta^{(N+1)}$ obeying the composition law
\begin{align}
\cosh\vartheta^{(N+1)}&=\cosh\vartheta\,\cosh\vartheta^{(N)}\nonumber\\
&\quad+\sinh\vartheta\,\sinh\vartheta^{(N)}\cos(\phi_N).
\end{align}
Here, $\phi_N$ is twice the phase sandwiched between the two matrices $\cT(\vartheta)$ and $\cT(\vartheta^{(N+1)})$, and is given in this case by $\phi_N=2(kL_N+\alpha+\beta^{(N)})$. We can regard $\phi_n$, in place of $L_n$, as the parameter that represents the disorder in our system.
Composition laws can also be written down for $\alpha^{(N+1)}$ and $\beta^{(N+1)}$, but they will not enter our analysis.

As a convenient shorthand, we will use the notation $C= \cosh\vartheta$, and $S=\sinh\vartheta=\sqrt{C^2-1}$. Similarly, one can define $C^{(N)}= \cosh\vartheta^{(N)}$ and $S^{(N)}= \sinh\vartheta^{(N)}$ so that the composition law appears succinctly as
\begin{equation}\label{eq:comp}
C^{(N+1)}=CC^{(N)}+SS^{(N)}\cos(\phi_N).
\end{equation}
The total transmission probability (or dimensionless conductance, in the language of transport on 1D wires) after $N$ scatterers is
\begin{equation}
\tau_N= \bigl(t^{(N)}\bigr)^2=\frac{2}{C^{(N)}+1}\,.
\end{equation}
The value of $\tau_N$ depends sensitively on the configuration of the scatterers, that is, on the values of the phases $\phi_n$.

\section{\label{sec:Recur}Recurrence relation: Averaging over disorder}
Rather than focusing on a particular configuration of the $N$ scatterers, one is usually more interested in statistics of the entire ensemble of configurations. 
This requires a statement about the nature of the disorder, namely, the probability distribution $\upd\mu(\phi_n)$ for $\phi_n$.
More generally, one can have correlated disorder, where the $\phi_n$ values are not statistically independent, but this is beyond the scope of our current discussion.
The transmission of $N$ scatterers, averaged over the disorder, is then
\begin{equation}
\langle\tau_N\rangle
=\int\upd\mu(\phi_{N-1})\ldots\int\upd\mu(\phi_2)\int\upd\mu(\phi_1)~\frac{2}{C^{(N)}+1}.
\end{equation}
The dependences on $\phi_1,\phi_2,\ldots,\phi_{N-1}$ are implicit in $C^{(N)}$. 
Writing out these dependences in full can be very complicated without being enlightening. 

To simplify the problem, we assume that the disorder is uniformly distributed, which permits the replacement
\begin{equation}
\int \upd\mu(\phi_n)\longrightarrow\int_{(2\pi)}\frac{\upd\phi_n}{2\pi}
\end{equation}
in the disorder average.
Here, the subscript $(2\pi)$ denotes integration over any $2\pi$-interval of our choice.
A uniform distribution is a good description whenever the separation between adjacent scatterers is itself uniformly distributed. 
It also describes the physical situation whenever $kL\gg 1$ for some typical separation $L$ between scatterers. 
This often applies for monochromatic light scattered by a stack of semi-transparent glass plates with layers of air between them.
In the case of a chain of ions trapped in a lattice, thermal motion of the ions within the trapping potential leads to random separation between adjacent ions. 
The separation is usually concentrated around some central value $L$, but for an electron incident with large momentum $k$, the phase $kL_n$ will explore many cycles of $2\pi$ over even small deviations of $L_n$ from $L$. 
A uniform distribution for $\phi_n\sim kL_n$ is then a fitting description.

\subsection{A recurrence relation}
To proceed with our analysis, we recall that the composition law \eqref{eq:comp} allows us to compute $\langle\tau_N\rangle$ recursively with the aid of the recurrence relation derived in Ref.~\onlinecite{Lu10}. 
We define a map $\cM_C$ that transforms any function $f(C')$ in accordance with
\begin{equation}
(\cM_C f)(C')= \int_{(2\pi)}\frac{\upd\phi}{2\pi}~f(CC'+SS'\cos\phi).
\end{equation}
Here, as before, $C= \cosh\vartheta$ is such that $t=\bigl[\cosh\bigl(\frac{1}{2}\vartheta\bigr)\bigr]^{-1}$ is the transmission amplitude for a single scatterer.
$C'$, the variable here, and $S'= \sqrt{C'^2-1}$ are to be viewed as the hyperbolic cosine and sine of some $\vartheta'$.
Note that $(\cM_C f)(C'=1)=f(C'=C)$.

The map $\cM_C$ possesses a symmetry that will prove useful later.
For any two functions $f(C')$ and $g(C')$,
\begin{equation}\label{eq:sym}
\int_1^\infty\upd C'(\cM_Cf)(C')\,g(C')=\int_1^\infty\upd C' f(C')\,(\cM_Cg)(C'),
\end{equation}
that is, we can consider $\cM_C$ as acting on either $f$ or $g$.
To see this, observe that
\begin{align}\label{eq:sym1}
&\quad~\int_1^\infty \upd C' (\cM_Cf)(C')g(C')\nonumber\\
&=\int_1^\infty \upd C'\int_1^\infty\upd C'' f(C')\,\cK_C(C',C'')\,g(C'')
\end{align}
with the kernel
\begin{align}
\cK_C(C',C'')&=\int_{(2\pi)}\frac{\upd\phi}{2\pi}\,\delta{\bigl(CC'+SS'\cos\phi-C''\bigr)}\nonumber\\
&=\frac{1}{\pi}\bigl[1-C^2-C'^2-C''^2+2C C' C''\bigr]^{-\frac{1}{2}}_+.\label{eq:K}
\end{align}
Here, $\delta(~)$ is the delta function, and the subscript $+$ is such that $[x]^{-1/2}_+=1/\sqrt{x}$ when $x\geq 0$, and is 0 otherwise. 
Since $\cK_C(C',C'')$ is invariant under permutations of $C,C'$ and $C''$, it follows that the roles of $f$ and $g$ in Eq.~\eqref{eq:sym1} can be interchanged, as expressed by the symmetry rule \eqref{eq:sym}.

From Ref.~\onlinecite{Lu10}, the average transmission probability is
\begin{equation}
\langle\tau_N\rangle=(\cM_C^{N-1}f)(C')\Big\vert_{C'=C}~~\textrm{for }f(C')= \frac{2}{C'+1},
\end{equation}
a fact that can be verified by writing out $\langle \tau_N\rangle$ in full for $N=1,2,3,\ldots$.
More generally, the disorder average of any function of $C^{(N)}$ is given by
\begin{equation}\label{eq:avgf}
{\left\langle f{\left(C^{(N)}\right)}\right\rangle}=(\cM_C^{N-1}f)(C')\Big\vert_{C'=C}.
\end{equation}

\begin{figure}
\centering
\includegraphics[width=0.6\columnwidth]{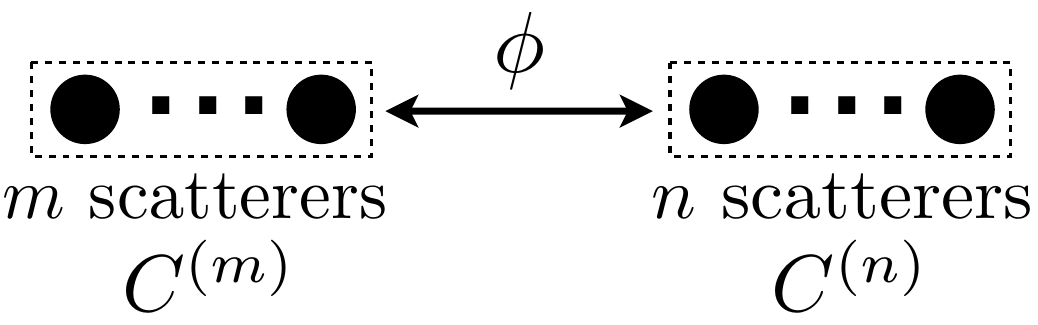}
\caption{\label{fig:Wmn} Illustration of concatenating two sub-chains of length $m$ and $n$ together to form a single chain of $N=m+n$ scatterers.}
\end{figure}

Another way of deriving Eq.~\eqref{eq:avgf} is to consider the probability density $W_N(C')$ that $C^{(N)}$ takes the value $C'$, so that
\begin{equation}\label{eq:avgfWN}
{\left\langle f{\left(C^{(N)}\right)}\right\rangle}=\int_1^\infty\upd C'\, f(C')\,W_N(C').
\end{equation}
In view of the composition law \eqref{eq:comp}, we have, for $N=m+n$, with $m,n$ positive integers,
\begin{widetext}
\begin{eqnarray}
W_N(C')&=&\int_1^\infty \upd C_1\, W_m(C_1)\int_1^\infty \upd C_2\, W_n(C_2)\int_{(2\pi)}\frac{\upd\phi}{2\pi}\,\delta{\bigl(C_1C_2+S_1S_2\cos\phi-C'\bigr)}\nonumber\\
&=&\int_1^\infty \upd C_1 \int_1^\infty \upd C_2  \,W_m(C_1) \,\cK_{C'}(C_1,C_2)\,W_n(C_2).\label{eq:WNnm}\label{eq:WNnm2}
\end{eqnarray}
\end{widetext}
The first line of Eq.~\eqref{eq:WNnm2} can be understood by splitting the $N$ scatterers into two segments (see Figure \ref{fig:Wmn}), one comprising the left $m$ scatterers, with an overall $C^{(m)}$ value equal to $C_1$, the other comprising the remaining $n$ scatterers, with an overall $C^{(n)}$ value equal to $C_2$. The two segments are separated by a random phase $\phi$, and the composition law gives the overall $C^{(N)}$ value for the $N$ scatterers.
For $W_1(C')=\delta(C'-C)$, Eq.~\eqref{eq:WNnm} gives iteratively $W_2(C'),W_3(C'), \ldots$, summarized as
\begin{equation}\label{eq:WNrecur}
W_N(C')=(\cM_C W_{N-1})(C')=(\cM_C^{N-1}W_1)(C').
\end{equation}
Upon inserting this formula into Eq.~\eqref{eq:avgfWN} and recalling the symmetry Eq.~\eqref{eq:sym} of $\cM_C$, we get Eq.~\eqref{eq:avgf} for the disorder average of $f$.

As an example of the usefulness of Eq.~\eqref{eq:avgf}, let us compute the average of $\log\tau_N$.
We set $f(C')=\log{\bigl(\frac{2}{C'+1}\bigr)}$. 
Performing the integration over $\phi$ in $(\cM_Cf)(C')$, we observe that $(\cM_Cf)(C')=f(C)+f(C')$. 
Repeated applications of $\cM_C$ thus yields
\begin{equation}\label{eq:logtau}
\langle\log\tau_N\rangle=Nf(C)=N\log\tau_1,
\end{equation}
where $\tau_1=t^2=\frac{2}{C+1}$ is the transmission probability for a single scatterer.
This expression comes as no surprise as $\log\tau_N$ is well known to be
additive under disorder averaging, and in the current context, the result
\eqref{eq:logtau} is the main observation in the paper by Berry and
Klein;\cite{Berry97}
Eq.~(\ref{eq:logtau}) can also be found in the paper by Perel' and
Polyakov \cite{PP84} as an unnumbered equation in \S4.

Note that formula \eqref{eq:avgf} applies even for disorder that is not uniformly distributed, as long as we replace $\int_{(2\pi)} \frac{\upd\phi}{2\pi}$ in the definition of $\cM_C$ by the more general $\int \upd\mu(\phi)$.
Equation \eqref{eq:WNnm} also holds for a general $W_1(C')$ not necessarily equal to a delta function.

\subsection{Eigenfunctions of the recurrence map}
Computing the average of $\log\tau_N$ is easy because $\cM_C$ acts  in a simple way on the relevant $f(C')$.
For more general functions, $\cM_C$ can act in a complicated manner, and it becomes difficult to solve the recurrence relation directly to obtain a closed-form expression for the averaged quantity.
The properties of the map $\cM_C$ thus require thorough understanding before we can proceed further.
Since we are to apply $\cM_C$ repeatedly, eigenfunctions of $\cM_C$---functions that remain invariant (apart from an overall factor) under the action of $\cM_C$---will be particularly useful.

Consider the Legendre functions, $\Lambda_\nu(C')$, which are functions that satisfy the Legendre differential equation (see, for example, Ref.~\onlinecite{AS}),
\begin{equation}
\frac{\upd}{\upd C'}(1-C'^2)\frac{\upd}{\upd C'}\Lambda_\nu(C')+\nu(\nu+1)\,\Lambda_\nu(C')=0.
\end{equation}
For our current purposes, as the notation already suggests, $C'$ is to be viewed as the hyperbolic cosine of some parameter, and we are interested in $C'$ between 1 and $\infty$. 
For a given $\nu$, there are two independent solutions to the Legendre equation, $P_\nu(C')$ (Legendre functions of the first kind) and $Q_\nu(C')$ (Legendre functions of the second kind). $Q_\nu(C')$ blows up at $C'=1$ while $P_\nu(C'=1)=1$ for all $\nu$.

Suppose we begin with $P_\nu(C')$ and apply the recurrence map $\cM_C$. 
We recall the addition theorem for $P_\nu(C')$ (see, for example, Ref.~\onlinecite{GR}),
\begin{align}
&\quad~ P_\nu(CC'+SS'\cos\phi)\\
&=P_\nu(C)P_\nu(C')+2\sum_{m=1}^\infty P_\nu^{-m}(C_<)P_\nu^m(C_>)\cos(m\phi)\nonumber
\end{align}
where $P_\nu^m$ are the associated Legendre functions, and $C_{<(>)}= \min(\max)\{C,C'\}$.
Employing this formula in the integrand of $(\cM_CP_\nu)(C')$ yields the eigenvalue equation
\begin{equation}\label{eq:evalEq}
(\cM_CP_\nu)(C')=P_\nu(C)P_\nu(C'),
\end{equation}
and the Legendre functions $P_\nu(C')$ are eigenfunctions of $\cM_C$. 
One can directly verify that $(\cM_CP_\nu)(C')$ satisfies the Legendre equation with the same value of $\nu$, which permits writing $(\cM_CP_\nu)(C')$ as a linear combination of $P_\nu(C')$ and $Q_\nu(C')$. 
Considering the value of $(\cM_CP_\nu)(C')$ at $C'=1$ leads to the conclusion \eqref{eq:evalEq}.
This eigenfunction property results in simple behavior of $P_\nu(C')$ under repeated applications of $\cM_C$,
\begin{equation}\label{eq:eigen}
(\cM_C^{n}P_\nu)(C')=P_\nu(C)^nP_\nu(C').
\end{equation}

The degree $\nu$ can be any complex number.
Of particular importance to us are the cases when $\nu$ is a nonnegative integer, and when $\nu$ takes the form $\nu=-\frac{1}{2}+\mathrm{i}x$. 
When $\nu=0,1,2,\ldots$, we have the Legendre polynomials familiar from many areas of physics;
the Legendre functions $P_{-\frac{1}{2}+\mathrm{i}x}(C')$ are known as the Mehler (or conical) functions, and have appeared in other physical problems, for example, the solution of Laplace's equation in toroidal coordinates. 
More relevant to our current subject, the Mehler functions were used in the exact solution of the DMPK equation (see Ref.~\onlinecite{Beenakker} and references therein), as well as in a different eigenvalue problem for the Anderson model.\cite{Abrikosov78,Kirkman84}

\section{\label{sec:Resistance}Moments of the resistance}
In the language of transport in electronic systems, the analogous quantity for transmission probability is the (dimensionless) conductance. 
The reciprocal of the conductance is the resistance, $\rho_N=1/\tau_N=\frac{1}{2}(C^{(N)}+1)$ for $N$ scatterers.
Our present concern is to compute the disorder-averaged resistance.
We begin with $f(C')=\frac{1}{2}(C'+1)=\frac{1}{2}[P_1(C')+P_0(C')]$, which gives
\begin{equation}\label{eq:24}
\langle\rho_N\rangle=\frac{1}{2}(C^N+1),
\end{equation}
upon applying Eq.~\eqref{eq:eigen}.\cite{Note:PP84-1}
For a long chain of scatterers, $N\gg 1$, we have $\log{\left\langle \rho_N\right\rangle}\simeq N\log C$, consistent with the usual expectation that the resistance grows exponentially as $N$ increases.

For averages of moments of the resistance, $\langle \rho_N^m\rangle$, for positive integer $m$, we start with \mbox{$f(C')=\big[\frac{1}{2}(C'+1)\big]^m$}. Noting that $f(C')$ involves only positive integer powers of $C'$, and every positive integer power of $C'$ can be expanded in terms of the Legendre polynomials $P_\ell(C')$ (see, for example, Ref.~\onlinecite{AS} or \onlinecite{GR}), we can similarly obtain $\langle \rho_N^m\rangle$ without great effort by applying Eq.~\eqref{eq:eigen}.

To illustrate, let us consider $\langle \rho_N^3\rangle$. 
Here, $f(C')=\frac{1}{8}(C'+1)^3=\frac{1}{20}{\left[P_3(C')+5P_2(C')+9P_1(C')+5P_0(C')\right]}$.
From Eq.~\eqref{eq:eigen}, we then get
\begin{equation}
\langle\rho_N^3\rangle
=\frac{1}{20}P_3(C)^N+\frac{1}{4}P_2(C)^N+\frac{9}{20}P_1(C)^N+\frac{1}{4},
\end{equation}
an expression that would have been difficult to obtain had we tried to solve the recurrence relation directly.
Note that the averages $\langle \rho_N^m\rangle$ can be computed iteratively starting with $m=1$ by employing recurrence formulas for Legendre polynomials that relate $C'P_\ell(C')$ to $P_{\ell\pm 1}(C')$.

For large $N$, $\langle\rho_N^m\rangle$ is dominated by the Legendre polynomial with the largest index $\ell = m$, and hence,
\begin{equation}
\langle\rho_N^m\rangle\simeq \frac{(m!)^2}{(2m)!}P_m(C)^N\quad\textrm{for }N\gg 1.
\end{equation}
Thus, $\log\langle \rho_N^m\rangle\propto N$ for $N\gg 1$ and any fixed $m$.
Also, $\rho_N^m$ has spread $\sqrt{\langle \rho_N^{2m}\rangle-\langle\rho_N^m\rangle^2}$, which, for large $N$, is dominated by $\sqrt{\langle \rho_N^{2m}\rangle}$.

\section{\label{sec:Conductance}Moments of the conductance}
To compute the average transmission probability (or conductance), we note that the above method of expanding powers of $C'$ in terms of Legendre polynomials no longer works, since the $C'$ dependence in $\tau_N$ occurs in the denominator. 
Whereas the Legendre polynomials are complete for $-1\leq C'\leq 1$, this completeness is of no help for $C'>1$, which is the range of interest in the current context.

Nevertheless, the idea of expanding in terms of Legendre functions still works.
For studying the conductance and its moments, we expand, not in terms of Legendre polynomials, but in terms of Legendre functions of the Mehler type $P_{-\frac{1}{2}+\mathrm{i}x}(C')$, through the Mehler-Fock transformation. 

\subsection{The Mehler-Fock transformation}
The Mehler-Fock transformation \cite{MehlerFock} is an index transformation  that uses the Mehler functions $P_{-\frac{1}{2}+\mathrm{i}x}(C')$ as the basis for expansion. Formally, one defines the Mehler-Fock transform $\hat f(x)$ of $f(C')$ as
\begin{equation}
\hat f(x)=x\tanh(\pi x)\int_1^\infty \upd C'\, P_{-\frac{1}{2}+\mathrm{i}x}(C')\,f(C').
\end{equation}
This integral exists whenever $f(C')$ is (weighted) square-integrable \cite{Yaku97} such that
\begin{equation}\label{eq:sqInt}
\int_1^\infty \upd C' \,\sqrt{C'-1}\,\big |f(C')\big|^2<\infty.
\end{equation}
The inverse transformation expresses the original function in terms of its transform,
\begin{equation}\label{eq:MFinv}
f(C')=\int_0^\infty \upd x \,P_{-\frac{1}{2}+\mathrm{i}x}(C')\,\hat f(x).
\end{equation}
The Mehler-Fock transform allows us to easily compute averages of physical quantities $f(C^{(N)})$, whenever $f$ is square integrable in the sense of \eqref{eq:sqInt}. We write,
\begin{align}
\langle f(C^{(N)})\rangle&=(\cM_C^{N-1}f)(C')\Big\vert_{C'=C}\nonumber\\
&={\left.\int_0^\infty \upd x\,(\cM_C^{N-1}P_{-\frac{1}{2}+\mathrm{i}x})(C')\,\hat f(x)\right\vert}_{C'=C}\nonumber\\
&=\int_0^\infty \upd x\,P_{-\frac{1}{2}+\mathrm{i}x}(C)^N\hat f(x),\label{eq:MFAvg}
\end{align}
where, exploiting the linearity of the map $\cM_C$, we applied $\cM_C$ to the Mehler functions using \eqref{eq:eigen}. This gives a closed-form expression for $\langle f(C^{(N)})\rangle$, with two remaining items to evaluate---the Mehler-Fock transform of $f$, and the final $x$ integral---but otherwise takes care of the complicated recursive composition law for concatenating $N$ scatterers.

\subsection{Moments of the conductance}
For moments of the transmission probability, or equivalently, moments of the conductance, the Mehler-Fock transform $\hat f(x)$ can be worked out explicitly. 
For $\langle\tau_N^m\rangle$, we set
\begin{equation}\label{eq:fMoments}
f(C')={\left(\frac{2}{C'+1}\right)}^m\qquad\textrm{for } m> \frac{1}{2}.
\end{equation}
To compute its Mehler-Fock transform, we note that $P_{-\frac{1}{2}+\mathrm{i}x}(C')$ can be written in terms of the hypergeome\-tric function,\cite{AS} $P_\nu(C')=\bigl(\frac{2}{C'+1}\bigr)^{-\nu}F\bigl(-\nu,-\nu;1;\frac{C'-1}{C'+1}\bigr)$ for $\nu=-\frac{1}{2}+\mathrm{i}x$.
Inserting this into the Mehler-Fock transform integral for $f$ gives
\begin{align}
\hat f(x)&=2x\tanh(\pi x)\\
&\quad\times \int_0^1\upd y\,{\left(1-y\right)}^{m-\frac{3}{2}-\mathrm{i}x}F{\left(\frac{1}{2}-\mathrm{i}x,\frac{1}{2}-\mathrm{i}x;1;y\right)},\nonumber
\end{align}
after making the substitution $y= \frac{C'-1}{C'+1}$.
This integral is a standard one,\cite{GR}
\begin{align}
\hat f(x)&=2x\tanh(\pi x)\,G_m(x)\\
\textrm{with }~G_m(x)&= {\left\vert\frac{\Gamma{\left(m-\frac{1}{2}+\mathrm{i}x\right)}}{\Gamma(m)}\right\vert}^2\nonumber\\
&=\frac{{\left(m-\frac{3}{2}\right)}^2+x^2}{(m-1)^2}\,G_{m-1}(x),\nonumber
\end{align}
where $\Gamma(z)$ is the familiar Gamma function.
For $m=1$, $\hat f(x)=2\pi x\tanh(\pi x)\,\sech(\pi x)$,
an expression that can be verified against known integrals involving Mehler functions (see, for example, Ref.~\onlinecite{Erdelyi}).

With this, we arrive at a compact expression for $\langle\tau_N^m\rangle$,
\begin{equation}\label{eq:taum}
\langle\tau_N^m\rangle=2\int_0^\infty \upd x\,x\tanh(\pi x)\,G_m(x)\,P_{-\frac{1}{2}+\mathrm{i}x}(C)^N,
\end{equation}
which is valid for $m>1/2$, including noninteger $m$ values.\cite{Note:PP84-2}
The remaining integral over $x$ can be estimated in limiting cases (for example, the weak-scattering limit discussed below), or evaluated numerically. 
We note that the Mehler functions are standard special functions for which efficient methods of numerical evaluation are known (for instance, via integral representations) or even built into mathematical software.

\section{\label{sec:WeakScat}A long chain of weak scatterers}
To connect with previous work on this subject, let us examine the limit of weak scattering, where we take $C$ close to 1. 
At the same time, we assume a long chain of scatterers, so that $N\gg 1$.

We consider $\langle \tau_N^m\rangle$, given by Eq.~\eqref{eq:taum}. 
For fixed $m$ and $C$, the integrand in Eq.~\eqref{eq:taum} is significant only for small $x$ values. 
This can be understood from the following integral representation for the Mehler function (see, for example, Ref.~\onlinecite{GR}),
\begin{equation}\label{eq:PIntRep}
P_{-\frac{1}{2}+\mathrm{i}x}(C)=\frac{\sqrt 2}{\pi}\int_0^{\vartheta} \upd u\,\frac{\cos(xu)}{\sqrt{C-\cosh u}},
\end{equation}
where $C=\cosh\vartheta$ as usual.
Since $x$ enters only in the cosine in Eq.~\eqref{eq:PIntRep}, for fixed $C$, $P_{-\frac{1}{2}+\mathrm{i}x}(C)$ stays bounded as a function of $x$.
Furthermore, one can check that the factor $x\tanh(\pi x)\,G_m(x)$ in the integrand of Eq.~\eqref{eq:taum} vanishes for large $x$ due to the suppression from the factorials in $G_m(x)$ as $x$ grows.
These observations justify the approximation that, for $C$ close enough to 1, or equivalently, for $\vartheta$ close enough to zero, one can expand the cosine in Eq.~\eqref{eq:PIntRep} about $xu=0$, and keep only the low-order terms,
\begin{align}
P_{-\frac{1}{2}+\mathrm{i}x}(C)&\simeq P_{-\frac{1}{2}}(C)+\frac{x^2}{2}\frac{\partial^2}{\partial x^2}P_{-\frac{1}{2}+\mathrm{i}x}(C)\Big\vert_{x=0}\nonumber\\
&\simeq P_{-\frac{1}{2}}(C)\mathrm{e}^{-bx^2}.\label{eq:Gauss}
\end{align}
In the second line, we have defined $b>0$ such that
\begin{equation}
b= -\frac{1}{2}[P_{-\frac{1}{2}}(C)]^{-1}\frac{\partial^2}{\partial x^2}P_{-\frac{1}{2}+\mathrm{i}x}(C)\Big\vert_{x=0},
\end{equation}
and approximated $1-bx^2\simeq \mathrm{e}^{-bx^2}$.

With this, $\langle\tau_N^m\rangle$ becomes much simpler: 
\begin{align}
\langle\tau_N^m\rangle&=2[P_{-\frac{1}{2}}(C)]^N\cI_m(x),\nonumber\\
\textrm{with}\quad \cI_m(x)&=\int_0^\infty \upd x \,x\tanh(\pi x)\,G_m(x)\,\mathrm{e}^{-Nbx^2}.
\end{align}
For $Nb\gg1$, the Gaussian suppression in $\cI_m(x)$ allows one to approximate the integral by Taylor-expanding the integrand about $x=0$. 
This gives
\begin{equation}
\cI_m(x)\simeq \frac{1}{4}{\left(\frac{\pi}{Nb}\right)}^{3/2}{\Big[\frac{\Gamma{\left(m-\frac{1}{2}\right)}}{\Gamma(m)}\Big]}^2, 
\end{equation}
with equality attained when $Nb\rightarrow \infty$.
Putting the pieces together, we have
\begin{eqnarray}
\langle\tau_N^m\rangle&\simeq&{\left[\frac{\Gamma{\left(m-\frac{1}{2}\right)}}{\Gamma(m)}\right]}^2\frac{1}{2}{\left(\frac{\pi}{Nb}\right)}^{3/2}[P_{-\frac{1}{2}}(C)]^N,~\label{eq:taumLimit}
\end{eqnarray}
valid in the limit of a long chain of weak scatterers.
This expression validates the conjecture of Ref.~\onlinecite{Lu10}, namely,
\begin{equation}
{\left(\frac{\langle\tau_N\rangle}{\langle\tau_2\rangle}\right)}^{1/(N-2)}\longrightarrow P_{-\frac{1}{2}}(C) \qquad\textrm{as }N\rightarrow\infty,
\end{equation}
since the quantity \mbox{$\Upsilon(\tau_1)=\int_{(2\pi)}\frac{\upd\varphi}{2\pi}(C+S\cos\varphi)^{-1/2}$} in Ref.~\onlinecite{Lu10} is an integral representation of $P_{-\frac{1}{2}}(C)$.\cite{GR}

We can further approximate $b$ and $P_{-\frac{1}{2}}(C)$ for $\vartheta\ll1$,
\begin{align}
P_{-\frac{1}{2}}(C)&=1-\frac{1}{16}\vartheta^2+O(\vartheta^4)\simeq \mathrm{e}^{-\vartheta^2/16}\nonumber\\
\quad\textrm{and}\quad
b&=\frac{1}{4}\vartheta^2+O(\vartheta^4).
\end{align}
From these, $P_{-\frac{1}{2}}(C)\simeq \mathrm{e}^{-b/4}$, which upon inserting into Eq.~\eqref{eq:taumLimit}, gives
\begin{equation}\label{eq:taumLimit1}
\langle\tau_N^m\rangle\simeq{\left[\frac{\Gamma{\left(m-\frac{1}{2}\right)}}{\Gamma(m)}\right]}^2\frac{1}{2}{\left(\frac{\pi}{Nb}\right)}^{3/2}\mathrm{e}^{-Nb/4}.
\end{equation}
This expression is identical to that found in Ref.~\onlinecite{Abrikosov81} for the conduction of electrons in a 1D wire, once we make the identification $L/l= Nb$, where $L$ is the length of the wire, and $l$ is the mean-free-path of the electrons.
The factor of $1/4$ in the exponential in Eq.~\eqref{eq:taumLimit1} is the familiar ratio of $\log(\langle\tau_N\rangle)$ to $\langle\log\tau_N\rangle=N\log\tau_1$ (see, for example, Ref.~\onlinecite{Kramer93}).
From the expressions for $P_{-\frac{1}{2}}(C)$ and $b$ in the weak-scattering limit, one can compute corrections to the standard answer of $1/4$ as a function of $\vartheta$.

Observe from Eq.~\eqref{eq:taumLimit} that, under the approximations above, $\langle\tau_N^m\rangle\propto\langle\tau_N\rangle$, 
and that the proportionality factor depends only on $m$, but not on the scattering strength. 
We can derive this statement more directly, using the following identity,\cite{AS}
\begin{align}
G_m(x)&=\frac{1}{\pi}{\left[\frac{\Gamma{\left(m-\frac{1}{2}\right)}}{\Gamma(m)}\right]}^2\alpha_m(x)\,G_1(x)\nonumber\\
\textrm{with}\quad\alpha_m(x)&=\prod_{k=1}^{m-1}{\left(1+\frac{(2x)^2}{(2k-1)^2}\right)},
\end{align}
so that we can write (without any approximation),
\begin{align}
&\langle\tau_N^m\rangle=\frac{1}{\pi}{\left[\frac{\Gamma{\left(m-\frac{1}{2}\right)}}{\Gamma(m)}\right]}^2 \\
&\qquad\times{\Bigl\{2\int_0^\infty \upd x\,x\tanh(\pi x)\alpha_m(x)G_1(x)P_{-\frac{1}{2}+\mathrm{i}x}(C)^N\Bigr\}}.\nonumber
\end{align}
The expression within the curly braces is exactly $\langle\tau_N\rangle$, if we can replace $\alpha_m(x)$ by~1. 
Now, $\alpha_m(x)\simeq 1$ whenever $x\ll 1$, and this is a good approximation whenever the integral in the curly braces gets its dominant contribution from small $x$ values. 
From our analysis above for the limit of a long chain of weak scatterers, the Gaussian suppression from the $\mathrm{e}^{-Nbx^2}$ factor guarantees that, indeed, the integrand is important only for $x\ll 1$.

More generally, $P_{-\frac{1}{2}+\mathrm{i}x}(C)^N$ is significant for \mbox{$x\ll1$} only whenever \mbox{$(\sinh\vartheta/\vartheta)^{\frac{N}{2}}\gg1$} (see Appendix \ref{app:Mehler}). 
This condition reduces to $Nb\gg 1$ in the limit of weak scattering.
We thus conclude that
\begin{equation}
\langle\tau_N^m\rangle\simeq\frac{1}{\pi}{\left[\frac{\Gamma{\left(m-\frac{1}{2}\right)}}{\Gamma(m)}\right]}^2\langle\tau_N\rangle
\end{equation}
whenever $(\sinh\vartheta/\vartheta)^{\frac{N}{2}}\gg 1$, which holds when either $N$ or $\vartheta$ (or both) are large. 
This  applies beyond the limit of a long chain of weak scatterers. 
For example, one can numerically verify the excellent accuracy of the above approximation when $\vartheta=5$ and $N=5$, for which $(\sinh\vartheta/\vartheta)^{N/2}\simeq 850$, where neither are the scatterers weak, nor is the chain a long one.
This proportionality between moments of the conductance is a specific example of a similar statement applicable for more general types of disorder (but valid only when $N\rightarrow \infty$) previously pointed out in Ref.~\onlinecite{Pendry94}.

\section{\label{sec:Prob}The probability distributions for conductance and resistance}
Armed with all the positive integer moments $\langle\tau_N^m\rangle$, one expects to be able to reconstruct the probability distribution for the conductance for given $N$ and $C$. 
In Ref.~\onlinecite{Pendry94}, the probability distribution for the conductance was obtained by writing down an expansion of the Fourier transform of the distribution in terms of $\langle\tau_N^m\rangle$.
Approaching the probability distribution from a different angle, one can make use of our recurrence relation to propagate the initial single-scatterer distribution to the distribution for $N$ scatterers. 
Our approach uncovers features absent from a reconstruction via moments.

We already have an expression (Eq.~\eqref{eq:WNrecur}) for $W_N(C')$, which we repeat here,
\begin{align}
W_N(C')&=(\cM_C^{N-1}W_1)(C')\nonumber\\
&=(\cM_C^{N-1}\widetilde W_{1,C'})(C'')\Big\vert_{C''=C},\label{eq:WN2}
\end{align}
where, in the second equality, $\widetilde W_{1,C'}(C'')=\delta(C''-C')$, and we have made use of Eq.~\eqref{eq:sym}.
One can verify that $W_N(C')\geq 0$ for $1\leq C'<\infty$ [in fact, \mbox{$W_N(C')=0$} for $C'>\cosh(N\vartheta)$], and that $\int_1^\infty \upd C'\,W_N(C')=1$.
From Eq.~\eqref{eq:WN2}, we immediately have the probability distributions for the conductance $\tau'= \frac{2}{C'+1}$ and the resistance $\rho'=1/\tau'$, related by Jacobian transformations,
\begin{align}
&\quad~\upd C'\,W_N(C')\\
&=\upd\tau'\, \frac{2}{\tau'^2}\,(\cM_C^{N-1} W_1)(C')\Big\vert_{C'=\frac{2}{\tau'}-1},&0\leq\tau'\leq 1,\nonumber\\
&=\upd\rho'\, 2\,(\cM_C^{N-1}W_1)(C')\Big\vert_{C'=2\rho'-1},&1\leq \rho'<\infty.\nonumber
\end{align}

Observe that Eq.~\eqref{eq:WN2} bears resemblance with the expression for $\langle f(C^{(N)})\rangle$ in Eq.~\eqref{eq:MFAvg}, if we set $f(C'')=\widetilde W_{1,C'}(C'')$. 
It thus seems plausible that our previous method of the Mehler-Fock transform might aid us in simplifying the recurrence formula for $W_N(C')$.
The Mehler-Fock transform integral (over $C''$, with $C'$ held constant) gives
$\hat f(x)=x\tanh(\pi x)P_{-\frac{1}{2}+\mathrm{i}x}(C')$.
Forgetting for the moment that $f(C'')$ is not square-integrable in the sense of \eqref{eq:sqInt}, we employ the inverse-transform formula to yield
\begin{equation}\label{eq:WNsqInt}
W_N(C')=\int_0^\infty \upd x \,x \tanh(\pi x) \,P_{-\frac{1}{2}+\mathrm{i}x}(C)^N \,P_{-\frac{1}{2}+\mathrm{i}x}(C'),
\end{equation}
which is Eq.~(28) in Ref.~\onlinecite{PP84}.

For computing statistics of square-integrable functions of $C'$, we can safely regard Eq.~\eqref{eq:WNsqInt} as a true identity for distributions, since we recover Eq.~\eqref{eq:MFAvg} when we replace $W_N(C')$ in $\int_1^\infty\upd C'\, f(C')W_N(C')$ with the right-hand side of Eq.~\eqref{eq:WNsqInt}.
For computing the average of non-square-integrable functions of $C'$, like the moments of the resistance, the validity of Eq.~\eqref{eq:WNsqInt} is less clear.
If we take the long-chain ($s=Nb\rightarrow\infty$), weak-scattering limit ($\vartheta\ll 1$) of the right-hand side of Eq.~\eqref{eq:WNsqInt}, we obtain
\begin{align}
&\quad~\int_0^\infty \upd x\,x\tanh(\pi x)P_{-\frac{1}{2}+\mathrm{i}x}(C')P_{-\frac{1}{2}+\mathrm{i}x}(C)^N\nonumber\\
&\simeq\frac{s^{-3/2}\mathrm{e}^{-s/4}}{2\sqrt{2\pi}}\int_{\cosh^{-1}(C')}^\infty \upd u\frac{ue^{-u^2/(4s)}}{\sqrt{\cosh u-C'}}.\label{eq:WrongW}
\end{align}
Here, we have approximated $P_{-\frac{1}{2}+\mathrm{i}x}(C)^N\simeq e^{-s/4}e^{-sx^2}$ as before, and also made use of an integral representation for $P_{-\frac{1}{2}+\mathrm{i}x}(C')$ of the form \cite{Hobson}
\begin{equation}
P_{-\frac{1}{2}+\mathrm{i}x}(C')=\frac{\sqrt 2}{\pi}\coth(\pi x)\int_{\cosh^{-1}(C')}^\infty \upd u\,\frac{\sin(xu)}{\sqrt{\cosh u-C'}}.
\end{equation}
The limiting expression in Eq.~\eqref{eq:WrongW} is identical to the distribution for $C^{(N)}$ found in Ref.~\onlinecite{Gert59} by solving the DMPK equation, and also in Ref.~\onlinecite{Abrikosov81} for the resistance in the Anderson model.
This lends credibility to Eq.~\eqref{eq:WNsqInt}, even for moments of the resistance.
Furthermore, the fact that $f(C')=[(C'+1)/2]^m$ is not square-integrable is perhaps not troublesome because $W_N(C')=0$ for $C'>\cosh(N\vartheta)$ (although not apparent from Eq.~\eqref{eq:WNsqInt}), so that the $C'$ integral in $\langle f(C^{(N)})\rangle$ cuts off at $C'=\cosh(N\vartheta)$ rather than extending to infinity.

A different source of concern lies with how the singularity of the delta function in $W_1(C')$ propagates as $N$ grows. 
For $N=1$, the singularity occurs at $C'=C$;
for $N=2$, it occurs at $C'=1$, since $W_2(C') = \cK_C(C,C')$; 
for $N=3$, the singularity is at $C'=C$ again (see Appendix \ref{app:WN}); for $N=4$, it goes back to $C'=1$.
In fact, for any even $N$, we expect $W_N(C'=1)$ to be particularly large (or even infinite), because, as explained in detail in Appendix \ref{app:WN}, there exists an infinite family of phase configurations that attain $C'=1$ whenever $N$ is even. 
Since $W_N(C'=1)=W_{N-1}(C'=C)$, this large value of $W_N(C'=1)$ for $N$ even is inherited by $W_N(C'=C)$ for $N$ odd.
This again suggests that the large value of $W_N(C')$ occurs at values that are different for $N$ even and $N$ odd, putting into suspect the existence of an asymptotic probability distribution for large $N$.
Fortunately, using our exact expressions for $W_N(C')$, one can show that beyond $N=4$, $W_N(C')$ becomes finite everywhere, and the singularity present in small $N$ values goes away; see Appendix \ref{app:WN} for a detailed analysis.
This justifies the validity of the reconstruction of the probability distribution of the conductance via its moments in Ref.~\onlinecite{Pendry94}, which relies on a Fourier transform that comes with an ``equal almost everywhere" condition that automatically gets rid of singularities with zero measure.

\section{\label{sec:Conc}Summary}
We have shown how one can analytically derive the statistics of a 1D chain of identical scatterers of any length, with uniform phase disorder. 
Making use of the fact that Legendre functions are eigenfunctions of the recurrence relation governing the chain statistics, we obtained disorder-averaged moments of the resistance by expanding in terms of Legendre polynomials; disorder-averaged moments of the conductance came from expanding in terms of the Mehler functions via the Mehler-Fock transformation.
The probability distributions of the conductance and the resistance can also be written in terms of the recurrence relation, and we pointed out singularities absent in existing derivations.
These extra features, despite the fact that they play little role in the physics in the end, remind us of the importance of exact and analytical solutions, which may be the only way of identifying subtle properties and verifying the validity of common assumptions.

\begin{acknowledgments}
We thank Dmitry Polyakov for bringing Ref.~\onlinecite{PP84} to our attention.
This work is supported by the National Research Foundation and the Ministry of Education, Singapore. We would like to thank Christian Miniatura, Beno{\^i}t Gr{\'e}maud, Cord M{\"u}ller, and Lee Kean Loon for insightful comments.
\end{acknowledgments}

\appendix

\section{Mehler functions for large $x$}\label{app:Mehler}
We are interested in an approximate form for the Mehler function $P_{-\frac{1}{2}+\mathrm{i}x}(C)$ applicable for large $x$.
Following Ref.~\onlinecite{Chukhrukidze}, one can expand the Mehler function for large $x$ in terms of Bessel functions,
\begin{equation}
P_{-\frac{1}{2}+\mathrm{i}x}(\cosh\vartheta)=\frac{1}{\sqrt{\vartheta\sinh\vartheta}}\sum_{n=0}^\infty A_n(\vartheta)\frac{J_n(x\vartheta)}{x^n},
\end{equation}
where the first three of the coefficients $A_n(\vartheta)$ are
\begin{align}
A_0(\vartheta)&=\vartheta,\nonumber\\
A_1(\vartheta)&=\frac{\vartheta\coth\vartheta-1}{8},\nonumber\\
\textrm{and } \quad A_2(\vartheta)&=\frac{(3\vartheta\coth\vartheta+1)^2-8(2+\vartheta^2)}{64\vartheta}.
\end{align}
One can check that the $n=0$ term gives, by far, the largest contribution, so
\begin{equation}
P_{-\frac{1}{2}+\mathrm{i}x}(C)\simeq \sqrt{\frac{\vartheta}{\sinh\vartheta}}~J_0(x\vartheta)\qquad\textrm{for }x\gg 1.
\end{equation}

Let us consider $[P_{-\frac{1}{2}+\mathrm{i}x}(C)]^N$, in light of this approximate expression. Since $J_0(x\vartheta)\leq 1$ for all $x\vartheta$ values, we focus on the factor $(\vartheta/\sinh\vartheta)^{N/2}$. For $\vartheta\ll1$, we have $(\vartheta/\sinh\vartheta)^{N/2}\simeq\Big(1-\vartheta^2/12\Big)^N\simeq e^{-Nb/3}$,
upon recalling that $b\simeq\frac{1}{4} \vartheta^2$ when $\vartheta\ll 1$. This implies, for $\vartheta\ll 1$, that $[P_{-\frac{1}{2}+\mathrm{i}x}(C)]^N$ is small for large $x$ whenever $Nb\gg 1$. More generally, $[P_{-\frac{1}{2}+\mathrm{i}x}(C)]^N$ is small for large $x$ whenever
\begin{equation}
{\left(\frac{\sinh\vartheta}{\vartheta}\right)}^{N/2}\gg 1,
\end{equation}
which can happen either when $N\gg1$ or when $\vartheta$ is large.

\section{Singularities in $W_N(C')$}\label{app:WN}

Here, we examine in detail the argument outlined in Section \ref{sec:Prob} regarding singularities in $W_N(C')$.
We begin by studying how $W_N(C')$ behaves for small $N$ near $C'=1$ for $N$ even, and $C'=C$ for $N$ odd.
Since $W_1(C')=\delta(C'-C)$, it is singular at $C'=C$.
For $W_2(C')=(\cM_C W_1)(C')$, we have
\begin{equation}
W_2(C')=\cK_C(C,C')=\frac{1}{\pi}\bigl[(C'-1)(2C^2-C'-1)\bigr]^{-1/2},
\end{equation}
which is singular at $C'=1$.
Near the singularity, we have
\begin{equation}
W_2(C')\rightarrow\frac{1}{\pi\sqrt 2 S}(C'-1)^{-1/2}\qquad \textrm{for } C'\rightarrow 1.
\end{equation}
For $W_3(C')$, we first write it in terms of an elliptic integral,
\begin{align}
W_3(C')={\left\{
\begin{array}{ll}
\frac{2}{\pi^2 a_>} K{\bigl(\frac{a_<}{a_>}\bigr)}& \textrm{if }1\leq C'\leq 4C^3-3C\\
0&\textrm{otherwise}
\end{array}\right.},
\end{align}
where $a_{>(<)}=\max(\min)\bigl\{[(CC'+SS')-1][(C^2+S^2)-(CC'-SS')], 4S^3 S'\bigr\}$.
$K$ is the complete elliptic integral,\cite{AS}
\begin{align}
K(\kappa)&= \int_0^{\pi/2}\upd \alpha\,\frac{1}{\sqrt{1-\kappa^2(\sin\alpha)^2}}\nonumber\\
&=\frac{\pi/2}{(1-\kappa^2)^{1/4}}\,P_{-\frac{1}{2}}{\left(\frac{1-\frac{1}{2}\kappa^2}{1-\kappa^2}\right)},
\end{align}
which is finite everywhere for $0\leq \kappa\leq 1$ except at $\kappa=1$, corresponding to the point where $C'=C$.
Expanding $K(\kappa)$ about $\kappa=1$ (see, for example, Ref.~\onlinecite{GR}), we can extract the behavior near the singularity,
\begin{equation}\label{eq:W3sing}
W_3(C')\rightarrow -\frac{3}{2\pi^2 S^2}\ln|C'-C|~\quad \textrm{for }|C'-C|\rightarrow 0.
\end{equation}
$W_4(C')$ for $C'=1$ is also singular, with a behavior identical to that of $W_3(C')$ in the vicinity of $C'=C$.
Beyond $N=3$, it becomes difficult to obtain a closed-form expression for $W_N(C')$. 

Nevertheless, for $N$ even, we expect $W_N(C'=1)$ to be particularly large. This can be understood in the following way:
$C^{(N)}=C'=1$ corresponds to perfect transmission through all $N$ scatterers. For $N$ even, every configuration of phases $\phi_1,\ldots,\phi_{N-1}$ such that $\phi_{N/2}=\pi$, and $\phi_{N/2-j}+\phi_{N/2+j}=2\pi$, for $j=1, 2, \ldots, N/2-1$, gives $C'=1$. There are thus infinitely many perfect-transmission configurations, since only the sum of each pair is constrained, but not the individual phases (apart from the central one). Furthermore, $C'$ is stationary, as the $\phi_n$s vary, at $C'=1$. This stationary property, together with the infinite family of configurations attaining $C'=1$, will give a large, possibly infinite, value of $W_N$ at $C'=1$. This large value of $W_N(C'=1)$ for $N$ even is inherited by $W_N(C'=C)$ for $N$ odd because $W_N(C'=1)=W_{N-1}(C'=C)$.
This difference between $N$ even and $N$ odd in the location of the large, possibly singular value of $W_N$ puts the existence of an asymptotic probability distribution as $N\rightarrow\infty$ under suspicion.

Fortunately, one can show that beyond $N=4$, $W_N(C')$ becomes finite everywhere, and the singularity present for small $N$ values goes away.
The points of suspicion are at $C'=1$ for $N$ even and $C'=C$ for $N$ odd.
We note that we can write
\begin{equation}
W_{m+n}(C'=1)=\int_1^\infty dC''\, W_m(C'')\, W_n(C''),
\end{equation}
as can be shown using the symmetry property Eq.~\eqref{eq:sym} of $\cM_C$.
Then, we have
\begin{equation}
W_6(C'=1)=\int_1^\infty \upd C'' \,W_3(C'')^2.
\end{equation}
Since $W_3(C'')$ is singular at one point $C''=C$, the vici\-nity of that singularity dominates the integral, but the logarithmic singularity behavior from \eqref{eq:W3sing}, after squaring and integrating, becomes finite. 
This immediately also tells us that \mbox{$W_5(C'=C)=W_6(C'=1)$} is finite. 
Thereafter, for larger $N$, we have $W_N(C')$ finite everywhere.
We note that Perel' and Polyakov \cite{PP84} arrived at the same conclusions about singularities of the $W_m(C')$s.

\bibliographystyle{unsrt.bst}

\begin{thebibliography}{00}
\bibitem{Note:PP84-1}
An unnumbered equation in \S4 of Ref.~\onlinecite{PP84} is equivalent to Eq.~(\ref{eq:24}).
\bibitem{Note:PP84-2}
The $m=1$ case is an unnumbered equation in \S4 of Ref.~\onlinecite{PP84}.
\end{thebibliography}

\end{document}